\documentclass[sigconf]{acmart}

\AtBeginDocument{%
  \providecommand\BibTeX{{%
    \normalfont B\kern-0.5em{\scshape i\kern-0.25em b}\kern-0.8em\TeX}}}

\setcopyright{acmcopyright}
\copyrightyear{2018}
\acmYear{2018}
\acmDOI{10.1145/1122445.1122456}

\acmConference[Woodstock '18]{Woodstock '18: ACM Symposium on Neural
  Gaze Detection}{June 03--05, 2018}{Woodstock, NY}
\acmBooktitle{Woodstock '18: ACM Symposium on Neural Gaze Detection,
  June 03--05, 2018, Woodstock, NY}
\acmPrice{15.00}
\acmISBN{978-1-4503-XXXX-X/18/06}

\usepackage{amsmath}
\usepackage{amsfonts}
\usepackage{subfigure}

\usepackage{makecell}
\usepackage{multirow}

\begin{document}
\title{Learning Explicit User Interest Boundary for Recommendation}

\author{Jianhuan Zhuo}
\email{zhuojianhuan@iie.ac.cn}
\affiliation{Institute of Information Engineering, Chinese Academy of Sciences,Beijing,China \& School of Cyber Security, University of Chinese Academy of Sciences,Beijing,China}

\author{Qiannan Zhu}
\email{zhuqiannan@ruc.edu.cn}
\affiliation{Gaoling School of Artificial Intelligence, Renmin University of China, Beijing Key Laboratory of Big Data Management and Analysis Methods}

\author{Yinliang Yue}
\email{yueyinliang@iie.ac.cn}
\affiliation{Institute of Information Engineering, Chinese Academy of Sciences,Beijing,China \& School of Cyber Security, University of Chinese Academy of Sciences,Beijing,China}

\author{Yuhong Zhao}
\email{zhaoyuhong@iie.ac.cn}
\affiliation{Institute of Information Engineering, Chinese Academy of Sciences,Beijing,China}

\begin{abstract}
The core objective of modelling recommender systems from implicit feedback is to maximize the positive sample score $s_p$ and minimize the negative sample score $s_n$,
which can usually be summarized into two paradigms: the pointwise and the pairwise.
The pointwise approaches fit each sample with its label individually, which is flexible in weighting and sampling on instance-level but ignores the inherent ranking property.
By qualitatively minimizing the relative score $s_n - s_p$, the pairwise approaches capture the ranking of samples naturally but suffer from training efficiency. 
Additionally, both approaches are hard to explicitly provide a personalized decision boundary to determine if users are interested in items unseen.
To address those issues, we innovatively introduce an auxiliary score $b_u$ for each user to represent the User Interest Boundary(UIB) and individually penalize samples that cross the boundary with pairwise paradigms, i.e., the positive samples whose score is lower than $b_u$ and the negative samples whose score is higher than $b_u$.
In this way, our approach successfully achieves a hybrid loss of the pointwise and the pairwise to combine the advantages of both.
Analytically, we show that 
our approach can provide a personalized decision boundary and significantly improve the training efficiency without any special sampling strategy.
Extensive results show that our approach achieves significant improvements on not only the classical pointwise or pairwise models but also state-of-the-art models with complex loss function and complicated feature encoding.
\end{abstract}

	
\maketitle

\section{Introduction}
\label{sec:introduction}


\begin{figure}[h]
    \centering
    \includegraphics[width=0.97\columnwidth]{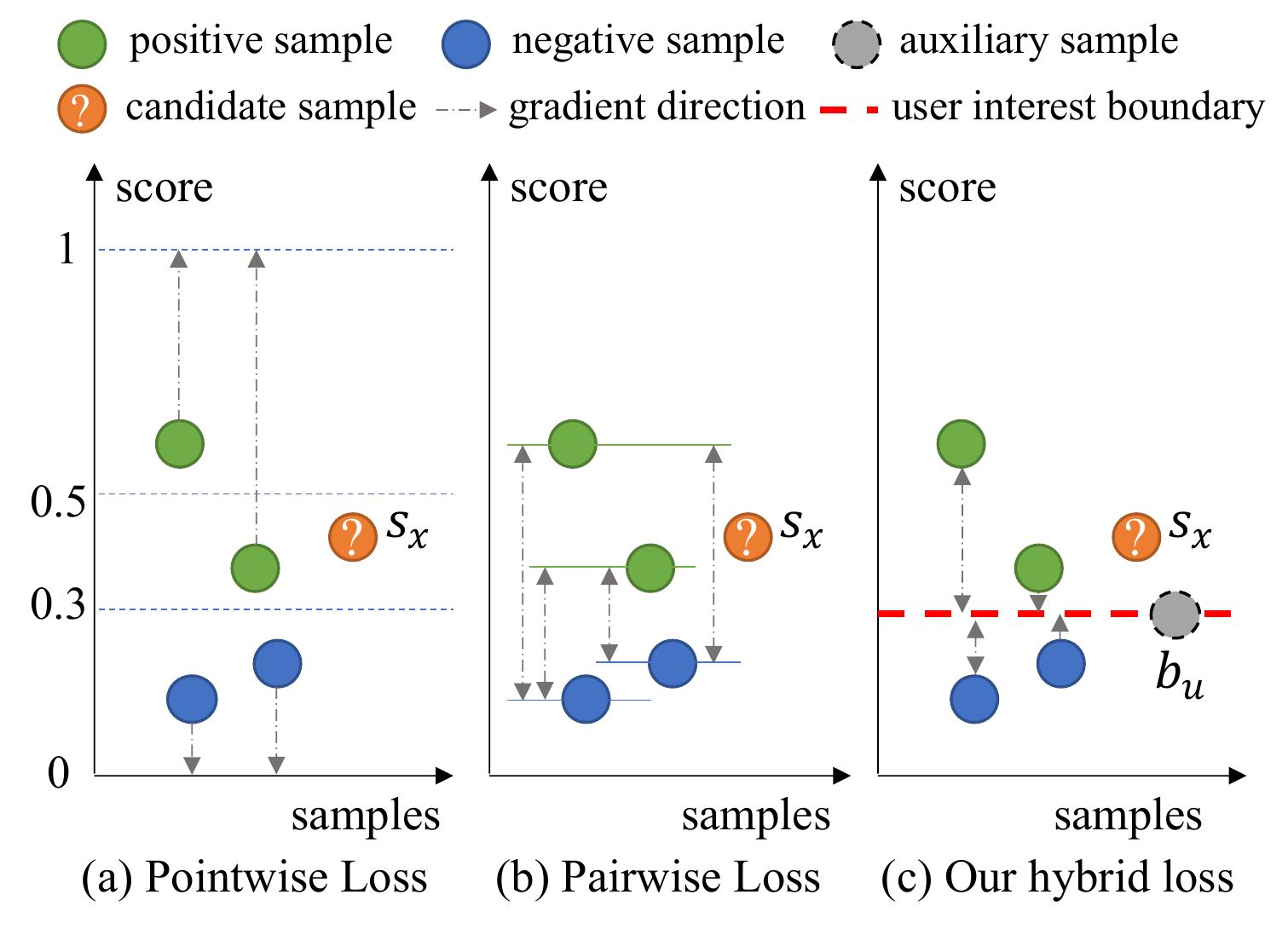}
    \caption{Loss paradigms comparison.}
    \Description{Loss paradigms comparison.}
    \label{fig:img-pattern}
\end{figure}

With the problem of information overload, 
recommendation system plays an important role to provide useful information for users efficiently.
As the widely used technique of the recommendation system, collaborative filtering (CF) based methods usually leverage the user interaction behaviours to model users' potential preferences and recommend items to users based on their preferences \cite{NCF}. 
Generally, given the user-item interaction data, a typical CF approach generally consists of two steps: (i) defining a scoring function to calculate the relevance scores between the user and candidate items, (ii) defining a loss function to optimize the total relevance scores of all observed user-item interaction.
From the view of the loss definition, the CF methods are usually optimized by a loss function that assigns higher scores $s_p$ to observed interactions (i.e., positive instances) and lower scores $s_n$ to unobserved interactions (i.e., negative instances). 

In the previous work, there are two types of loss functions designed for recommendation systems, namely, pointwise and pairwise.
As shown in figure~\ref{fig:img-pattern}(a),
pointwise based approaches typically formulate the ranking task as a regression or classification task, where the loss function $\psi(x, l)$ directly optimizes the normalized relevance score $s_x$ of the sample $x$ into its label $l \in \{0,1\}$. 
The sample $x = (u, i)$ is the observed or unobserved pair of the user $u$ and the item $i$. 
Usually, the pointwise-based loss function uses a fixed hardline like 0.5 as the indictor to distinguish the sample as positive or negative, i.e., for all users in the ranking stage, samples whose scores more than 0.5 are regarded as positive, where the user $u$ would be interested in the item $i$. 
Correspondingly, the pairwise based approaches shown in figure~\ref{fig:img-pattern}(b), take the pairs $(x_n, x_p)$ of the positive and negative samples as input, and try to minimize the relative score $s_n - s_p$ by the loss function $\phi(x_n, x_p)$. 
The pairwise loss focuses on making the score $s_p$ of positive samples greater than that $s_n$ of negative samples, which can get the plausibility of the sample $x = (u, i)$ for being used in the ranking stage.


Recently, such two paradigms of loss function are widely used in various recommendation methods, and helpful to achieve a promising recommendation performance. 
However, they still have disadvantages. 
They are hard to learn the \textbf{explicit user's personalized interest boundary}, which can directly infer whether the user likes a given unseen item ~\cite{boundary,YouTube}.
Factually, the users have their own interest boundary learned from their interactions for determining whether the sample is the positive sample in ranking stage.
As mentioned above, the pointwise loss function is non-personalized and prone to determine the global interest boundary as a fixed hardline for all users, which may make the incorrect classification for the users whose real interest boundary is lower than the fixed hardline. 
For example, in figure \ref{fig:img-pattern}(a), although the score $S_x$ of the sample $x$ is more than the users' real interest boundary, the positive sample $x$ is still classified into the negative group as its score $S_x$ is lower than the fix hardline. 
While the pairwise approach is unable to provide an explicit personalized boundary for unseen candidate sample $x$ in figure~\ref{fig:img-pattern}(b), because its score $S_x$ learned by the relative scores can only reflect the plausibility of the sample $x$, not the explicit user-specific indicator of whether the sample is a positive sample in ranking stage. 

In addition, we are also interested in another issue,
the \textbf{lower training efficiency} problem that prevents the pairwise models from obtaining optimal performance. 
For the pairwise, with the convergence of the model in the later stage of training, most of negative samples have been classified correctly. 
In this case, the loss of most randomly generated training samples is zero, i.e., those samples are too ``easy'' to be classified correctly and are unable to produce an effective gradient to update models, which is also called the gradient vanishing problem.
To alleviate this problem, previous work adopt the hard negative sample mining strategy to improve the probability of sampled effective instances~\cite{AdvIR,IRGAN,NSCaching,RNS-AS,SRNS}. 
Although these methods are successful, they ignore the basic mechanism leading to the vanishing problem. 
Intuitively, hard samples are typically those near the boundary, and it is hard for the model to distinguish these samples well.
Instead of improving the sampling strategy to mine a hard sample, 
why don't we directly use the boundary to represent the hard sample score?

To address the issue mentioned above, we innovatively introduce an auxiliary score $b_u$ for each user and individually penalize samples that cross the boundary with the pairwise paradigm, i.e., the positive samples whose score is lower than $b_u$ and the negative samples whose score is higher than $b_u$.
The boundary $b_u$ is meaningfully used to indicate user interest boundary(UIB), which can be used to explicitly determine whether an unseen item is worth recommending to users.
As we can see from the figure~\ref{fig:img-pattern}(c), the candidate sample $s_x$ can be easily predicted as positive by the UIB $b_u$ of user $u$.
In this way, our approach successfully achieves a hybrid loss of the pointwise and the pairwise to combine the advantages of both.
Specifically, it follows the pointwise in the whole loss expression while the pairwise inside each example.
In the ideal state, the positive and negative samples should be separated by the boundary $b_u$, i.e., scores of positive samples are higher than $b_u$ and those of negative are lower as figure~\ref{fig:img-pattern}. The learned boundary can provide a personalized interest boundary for each user to predict unseen items.
Additionally, different from previous works trying to mine hard samples, the boundary $b_u$ can be directly interpreted as score of the hard samples, which significantly improves the training efficiency without any special sampling strategy.
Extensive results show that our approach achieves significant improvements on not only classical models of the pointwise or pairwise approaches, but also state-of-the-art models with complex loss function and complicated feature encoding.

The main contributions of this paper can be summarized as:

\begin{itemize}
    \item We propose a novel hybrid loss to combine and complement the
     pointwise and pairwise approaches. As far as we know, it is the first attempt to introduce an auxiliary score for fusing pointwise and pairwise.
    
    \item We provide a efficient and effective boundary to match users' interest scope. As the result, a personalized interest boundary of each user is learned, which can be used in the pre-ranking stage to filter out abundant obviously worthless items.
    \item We analyze qualitatively the reason why the pairwise trains inefficiently, and further alleviate the gradient vanishing problem from the source by introducing an auxiliary score to represent as the hard sample.
    \item We conduct extensive experiments on four publicly available datasets to demonstrate the effectiveness of UIB, where multiple baseline models achieve significant performance improvements with UIB.
\end{itemize}

\section{Preliminary}
\label{sec:preliminary}
For the implicit CF methods, the user-item interactions are the important resource in boosting the development of recommender systems.
For convenience, we use the following consistent notations throughout this paper: 
the user set is $\mathcal{U}$ and the item set is $\mathcal{X}$.
Their possible interaction set is donoted as $\mathcal{T} =\mathcal{U} \times \mathcal{X}$, in which the observed part is regarded as users' real interaction histories $\mathcal{I} \subset \mathcal{T}$.
Formally, the labeling function $l: \mathcal{T} \to \{0, 1\}$ is used to indicate if a sample is observed, where a value of 1 denotes the interaction is positive (i.e. $(u,p) \in \mathcal{I}$) and a value of 0 denotes the interaction is negative (i.e. $(u,n) \notin \mathcal{I}$).

In implicit collaborative filtering, the goal of models is to learn a score function $s: \mathcal{T} \to \mathbb{R}$ to reflect the relevance between items and users.
The loss function is used to indicate how well the score function fits, which can usually be summarized into two paradigms: the pointwise and the pairwise.


\subsection{Pointwise loss}
The pointwise approach formulates the task as a classification task of single sample, whose loss function $\psi$ directly optimizes the normalized relevance score $s(u,x)$ between user $u$ and item $x$ into its label $l(u,x)$:
\begin{equation}
    \mathcal{L} = \sum_{(u,x) \in \mathcal{T}} \psi  \left(s(u,x), l(u,x) \right)
    \label{eq:pointwise}
\end{equation}
where $\psi$ can be CrossEntropy~\cite{NCF} or Mean-Sum-Error~\cite{ENMF} and so on.
Since the pointwise approach optimizes each sample individually, it is flexible on sampling and weighting on instance-level.
However, as these scores depend on the observation context, fitting the sample into the fixed score is hard to reflect the inherent ranking property~\cite{A-Case-Study}.

\subsection{Pairwise loss}
Instead of fitting samples with fixed scores, the pairwise approach tries to assign the positive samples with higher scores than that of the negative~\cite{burges2005learning}. The loss function of the pairwise approach can be rewritten as:
\begin{equation}
    \mathcal{L} = \sum_{(u,p) \in \mathcal{I}} \sum_{(u,n) \notin \mathcal{I}} \phi \left(s(u,n) - s(u,p) \right)
    \label{eq:pairwise}
\end{equation}
where $\phi$ can be MarginLoss~\cite{CML,SML} or LnSigmoid~\cite{NCF,LightGCN} and so on.
Although the pairwise approach can improve generalization performance by learning the qualitative scores between the positive and negative samples, it is hard to provide effective ranking information in the inference stage and suffers from the gradient vanishing problem. 

\begin{table}
    \begin{center}
    \begin{tabular}{ l | c | c | c }
    \hline
                        & Pointwise & Pairwise  & Ours \\
    \hline
    learn ranking       & $\times$  & $\surd$   & $\surd$ \\
    flexible sampling   & $\surd$   & $\times$  & $\surd$ \\
    flexible weighting  & $\surd$   & $\times$  & $\surd$ \\
    personalized boundary   & $\times$  & $\times$  & $\surd$ \\
    \hline                          
    \end{tabular}
\end{center}
\caption{Features comparison among loss paradigms.}
\label{table:features}
\end{table}

The pointwise and the pairwise approaches have their two sides as shown in table~\ref{table:features}.
Intuitively, a better loss function can be obtained by fully combining the advantages of the two to further improve the recommendation performance.
Hence, we seek a hybrid loss to combine and complement each other.


\section{Methodology}
\label{sec:method}



\subsection{A hybird loss}
We propose a new loss paradigm to combine the advantages of two mainstream methods and match the user's interest adaptively.
Our approach is effective and efficient.
As shown in equation~\ref{eq:mine}, we innovatively introduce an auxiliary score $b_u \in \mathbb{R}^1$ for each user $u$ to represent the User Interest Boundary(UIB):
\begin{equation}
    b_u = W^{\top}\mathbf{P}_u
    \label{eq:bu}
\end{equation}
where $\mathbf{P}_u \in \mathbb{R}^{d}$ is the embedding vector of user $u$, $W \in \mathbb{R}^{d}$ is a learnable vector.
Our loss comes from the weighted sum of two parts: the positive sample loss part $L_p$ and the negative sample loss part $L_n$. 
Within the $L_p$ and $L_n$, the pairwise loss is used to penalize samples that cross the decision boundary $b_u$, i.e. the positive samples whose $s(u,p)$ is lower than $b_u$ and the negative samples whose $s(u,n)$ is higher than $b_u$.
Formally, the loss function of our approach can be rewritten as:
\begin{equation}
    \mathcal{L} =   \underbrace{        \sum_{(u,p) \in     \mathcal{I}} \phi(b_u - s(u,p))}_{L_p} + 
                    \underbrace{\alpha  \sum_{(u,n) \notin  \mathcal{I}} \phi(s(u,n) - b_u)}_{L_n}
    \label{eq:mine}
\end{equation}
where $\alpha$ is a hyperparameter to balance the contribution weights of positive and negative samples.

Our approach can be regarded as \textbf{the hybrid loss of the pointwise and the pairwise}, which is significantly different from previous works.
On one hand, the pointwise loss usually optimizes each sample to match its label, which is flexible but not suitable for rank-related tasks.
On another hand, the pairwise loss takes a pair of the positive and negative samples, then optimizes the model to make their scores orderly, which is a great success but suffers from the gradient vanishing problem.
Our approach successfully combines and complements each other by introducing an auxiliary score $b_u$.
Specifically, in the whole loss expression, it follows the pointwise loss because the positive samples and negative samples are calculated separately in $L_p$ and $L_n$.
Inside the $L_p$ and $L_n$, 
each sample is applied the pairwise loss, e.g. the margin loss, with the auxiliary score $b_u$.
In other words, the pairwise loss is applied on $(b_u - s(u,p))$ and $(s(u,n) - b_u)$ respectively rather than traditional $(s(u,n) - s(u,p))$.
In this way, our approach can provide a flexible and efficient loss function.


\subsection{User Interest Boundary}
\label{sec:UIB}
The learned $b_u$ can provide a personalized decision boundary to determine whether the user likes the item in the inference stage.
The explicit boundary is useful for many applications, e.g. filter out abundant obviously worthless items in the pre-ranking stage.

\begin{figure}[h]
    \centering
    \includegraphics[width=0.7\columnwidth]{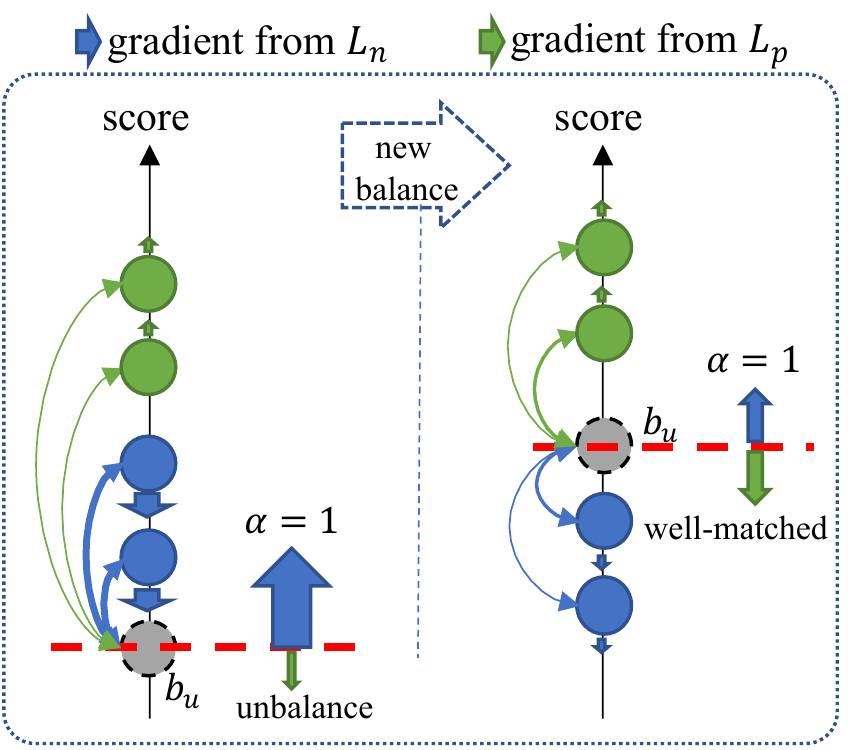}
    \caption{Adpatively matching of boundary}
    \label{fig:adpative-matching}
\end{figure}

Why it is personalized? From the view of gradient direction, the ideal boundary of user $u$ is adaptively matched under the balance between positive and negative samples, which provides a personalized decision boundary for different users. 
Concretely, to optimize the loss function equation~\ref{eq:mine} we proposed, the optimizer has to penalize two-part simultaneously: the positive part $L_p$ and the negative part $L_n$.
During the process, the $L_p$ upward forces the scores of positive samples and downward forces the boundary $b_u$ like the green arrow in figure~\ref{fig:adpative-matching}, while the $L_n$ upward forces the boundary $b_u$ and downward forces the negative like the blue arrow.
If the boundary is straying from its reasonable range, the unbalance gradient will push it back until it well-matched between positive and negative.
Take the boundary $b_u$ in figure~\ref{fig:adpative-matching} for example. The boundary is mistakenly initialized to a very low score so that all negative samples are incorrectly classified as positive. 
Consequently, the boundary is pushed upward to balance the low $L_p$ and large $L_n$.
As the result of the balance between positive and negative samples, the boundary can adaptively match the user's interest scope.
Additionally, we also conduct experiments to verify this statement, detailed in section~\ref{sec:uib}.

As our approach explicitly learns the user interest boundary in the training stage,
the boundary learned can be directly used to determine whether the user likes the item in the inference stage.
It is easily used, such as regarding candidates with a higher score than boundary as the positive, otherwise the negative as shown in figure~\ref{fig:img-pattern}(c).







\subsection{Training Efficiency}
\label{sec::training_eff}
\begin{figure}[h]
    \centering
    \includegraphics[width=0.98\columnwidth]{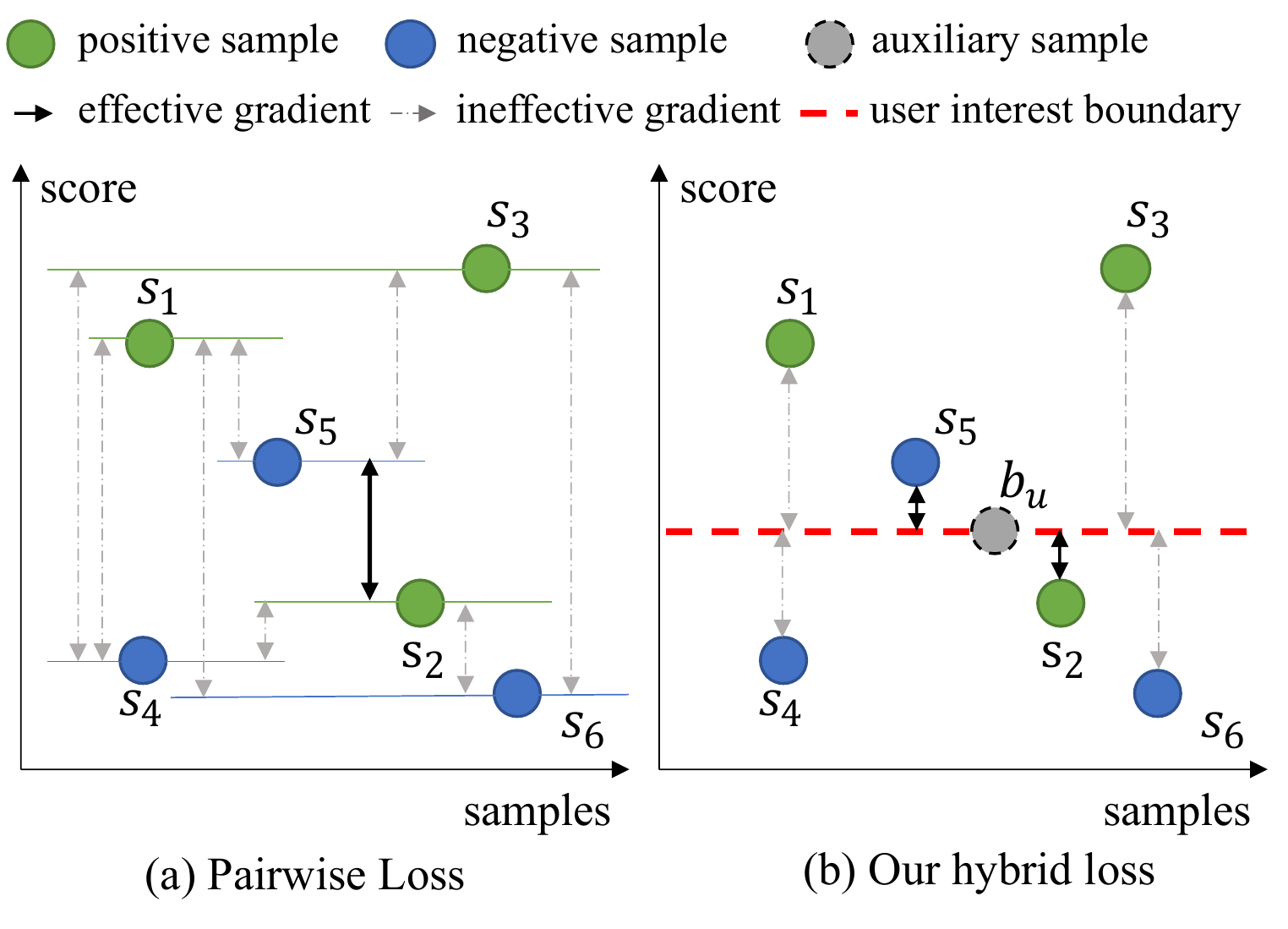}
    \caption{Training efficiency comparison.}
    \label{fig:bu_vash}
\end{figure}

Our approach can significantly improve the training efficiency without any special sampling strategy.
The traditional pairwise loss function suffers from the gradient vanishing problem, especially in the later stage of training.
Since the pairwise loss is to minimize the relative score $\Delta=(s_n-s_p)$, it is reasonable to use $\Delta$ as an indicator to explain the training efficiency.
For example, as shown in figure~\ref{fig:bu_vash}, 9 pair training instances are generated by combining the scores of positive examples \{$s_1$,$s_2$,$s_3$\} and the scores of negative examples \{$s_4$,$s_5$,$s_6$\}, but only the pair ($s_2$, $s_5$) can provide effective gradient information to update the model, i.e. is not classified correctly or ``corrupted''~\cite{Corupted} as $\Delta=(s_5-s_2)$ greater than zero. 
That is $1/9$ probability.
By introducing the boundary setting, our approach significantly improves the training efficiency only with the simple uniform sampling strategy.
From the view of negative sampling, the boundary $b_u$ can be naturally interpreted as the hard negative sample for positive samples and the hard positive for negative ones.
Concretely, both positive and negative samples are paired with the boundary $b_u$ and result in 6 pair training instances, two of which are effective, i.e., $(s_5-b_u)$ and $(b_u-s_2)$. 
That is $1/3$ probability.
Formally, let a dataset contains $N$ positive samples, $N$ negative samples, and $M$ effective pairs of all possible combination results. 
Each time a training instance $(s_p, s_n)$ is generated by randomly sampling, only $M/N^2$ probability of the pairwise loss can provides effective gradient information.
While in our approach, $M/N$ probability can be achieved.
Hence, compared with the traditional pairwise loss, our approach is more efficient and significantly alleviate the gradient vanishing problem. 
Furthermore, the advantage of our approach is proved experimentally in section~\ref{sec:analysis-efficiency}. 

\subsection{Classes balance}
\label{sec:class-balance}
\begin{figure}[h]
    \centering
    \includegraphics[width=0.8\columnwidth]{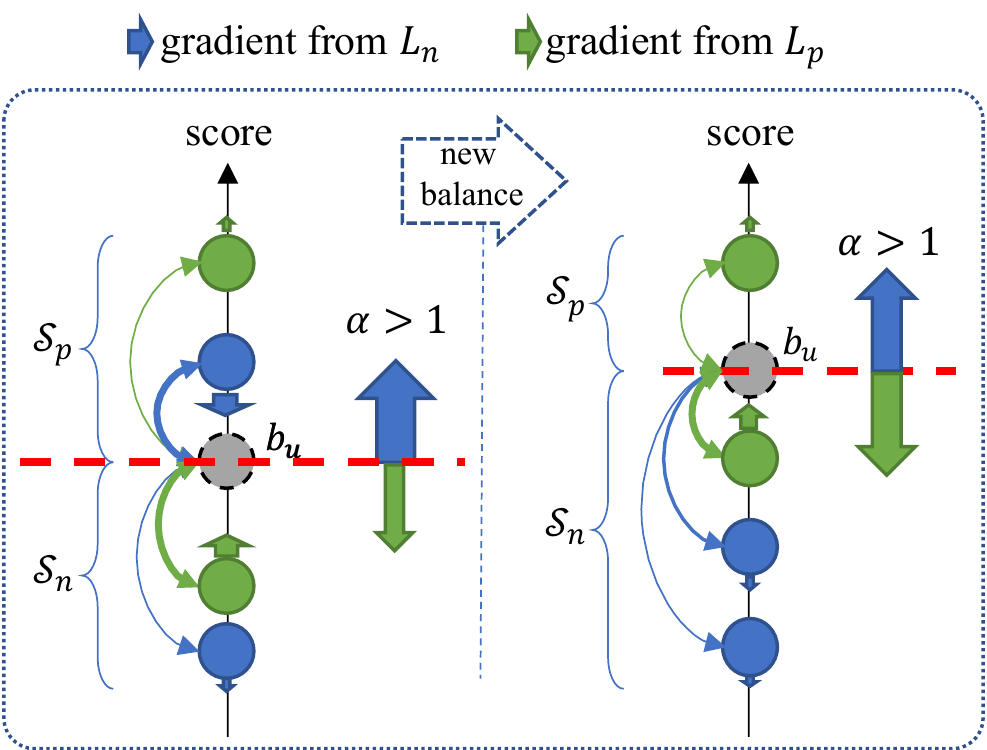}
    \caption{Boundary rebalance with larger $\alpha$ setting.}
    \label{fig:bu_rebalance}
\end{figure}

Our approach can provide a flexible way to balance the negative and positive samples. 
In general, the positive samples are observed instances collected from users' interaction history, while negative samples are instances that are not in the interaction history.
As the method is different to obtain the positive and negative samples, it is unreasonable to treat both with the same weight and sampling strategy.
As our approach optimizes $L_n$ and $L_p$ individually on the whole expression, we can assign a proper $\alpha$ and develop different sampling strategy to balance classes. 
For sampling strategy, since the negative sample space is much larger than the positive, we use negative samples of $M$ times the positive to balance the class for each batch sampling. 
Significantly, here we still use the simplest uniform sampling strategy instead of other advanced methods~\cite{AdvIR,IRGAN,NSCaching,RNS-AS,SRNS}.
For weighting, the introduction of the auxiliary score mechanism makes another way possible: adjusting $\alpha$ to enlarge or reduce the scoring space of positive and negative samples.
Here, we denote the positive scoring space $\mathcal{S}_p$ as the range between boundary and the maximum score of positive samples, while the negative scoring space $\mathcal{S}_n$ as the range between boundary and the minimum score of negative samples as shown in figure~\ref{fig:bu_rebalance}.
Intuitively, the scoring space corresponds to the score range of a candidate sample determined as the positive sample or negative.
Amplifying $\mathcal{L}_n$ by increasing $\alpha$ will push the boundary upward and compress the positive scoring space $\mathcal{S}_p$ as shown on the right side of figure~\ref{fig:bu_rebalance}.
With the boundary $b_u$ driven up, all positive samples obtain a larger upward gradient by larger $\mathcal{L}_p$ and gather into a more compact space.
In this way, the expected score of positive and negative samples are limited in a proper range. our approach can adjust $\alpha$ to balance the negative and positive samples.
Conversely, the same is true for reducing $\alpha$.
This phenomenon is also observed in the experiments with different $\alpha$ settings in section~\ref{sec:alpha}.

\section{Experiment}
\label{section:exp}
In this section, we first introduce the baselines (including the boosted version via our approach), datasets, evaluation protocols, and detailed experimental settings. 
Further, we show the experimental results and some analysis of them. In particular, our experiments mainly answer the following research questions:

\begin{itemize}
    \item \textbf{RQ1} How well can the proposed UIB boost the existing models?
    \item \textbf{RQ2} How well does the boundary match the users' interest scope?
    \item \textbf{RQ3} How does the $\alpha$ affect the behaviour of the boundary?
    \item \textbf{RQ4} How does the proposed UIB improve the training efficiency?
\end{itemize}

\subsection{Baselines}
To verify the generality and effectiveness of our approach, targeted experiments are conducted. 
Concretely, we reproduce the following four types of S.O.T.A. models as the baselines and implement the boosted version by our UIB loss. 
To compare the difference among those architectures, table~\ref{table:baselines} list all score function and loss function used in the baselines and our boosted models.

\begin{table*}[h]
    \begin{center}
    \begin{tabular}{ l | c | c }
    \hline
    \textbf{Method} &
    \textbf{Score} $s(u,x)$ &
    \textbf{Loss} $\mathcal{L}$ \\
    \hline
    BPR~\cite{BPR}  & \multirow{2}{*}{$\mathbf{P}^{\top}_u \mathbf{Q}_x$} 
                    & $ - \sum_{(u,p) \in \mathcal{I}} \sum_{(u,n) \notin \mathcal{I}} \text{ln}\sigma \left(s(u,p) - s(u,n)\right)$ \\ 
                    \cline{1-1} \cline{3-3} 
    BPR+UIB(ours)   &
                    & $ - \sum_{(u,p) \in \mathcal{I}} \text{ln}\sigma (s(u,p) - b_u) - \alpha \sum_{(u,n) \notin \mathcal{I}} \text{ln}\sigma (b_u - s(u,n))$ \\
    \hline
    NCF~\cite{NCF}  & \multirow{2}{*}{$f(\mathbf{P}_u, \mathbf{Q}_x)$} 
                    & $ - \sum_{(u,x)} l(u,x)\text{ln}(s(u,p)) + (1-l(u,x))\text{ln}(1-s(u,n))$\\ 
                    \cline{1-1} \cline{3-3} 
    NCF+UIB(ours)   & 
                    & $ - \sum_{(u,p) \in \mathcal{I}} \text{ln}\sigma (s(u,p) - b_u) - \alpha \sum_{(u,n) \notin \mathcal{I}} \text{ln}\sigma (b_u - s(u,n))$ \\
    \hline
    SML~\cite{SML}  & \multirow{2}{*}{$||\mathbf{P}_u - \mathbf{Q}_x||^2_2$} 
                    & \makecell[c]{$\sum_{(u,p) \in \mathcal{I}} \sum_{(u,n) \notin \mathcal{I}} |s(u,n) - s(u,p)+m_u|_+$ \\
                                   $+ \lambda |s(n,p) - s(u,p)+n_u|_+ \gamma \mathcal{L}$ } \\
                    \cline{1-1} \cline{3-3} 
    SML+UIB(ours)   &
                    & \makecell[c]{$\sum_{(u,p) \in \mathcal{I}} \sum_{(u,n) \notin \mathcal{I}} |s(u,n) - b_u + m_u|_+ $ \\ 
                                   $+ \alpha |b_u - s(u,p) + m_u|_+ + \lambda |s(n,p) - s(u,p)+n_u|_+ \gamma \mathcal{L}$} \\
    \hline
    LightGCN~\cite{LightGCN}  & \multirow{2}{*}{$\text{G}(u)^{\top} \text{G}(x)$}
                    & $ - \sum_{(u,p) \in \mathcal{I}} \sum_{(u,n) \notin \mathcal{I}} \text{ln}\sigma \left(s(u,p) - s(u,n)\right)$ \\
                    \cline{1-1} \cline{3-3} 
    LightGCN+UIB(ours)   & 
                    & $ - \sum_{(u,p) \in \mathcal{I}} \text{ln}\sigma (s(u,p) - b_u) - \alpha \sum_{(u,n) \notin \mathcal{I}} \text{ln}\sigma (b_u - s(u,n))$ \\
    \hline
    \end{tabular}
\end{center}
\caption{Architecture comparison among baselines and our boosted models.}
\label{table:baselines}
\end{table*}

\begin{itemize}
    \item \textbf{Pairwise model}, i.e. the BPR~\cite{BPR}, which applies an inner product on the latent features of users and items and uses a pairwise loss, the LnSigmoid(also known as the soft hinge loss) to optimize the model as follows:
    \begin{equation}
        \mathcal{L} =    - \sum_{(u,p) \in \mathcal{I}} \sum_{(u,n) \notin \mathcal{I}} \text{ln}\sigma \left(s(u,p) - s(u,n)\right).
        \label{eq:lns}
    \end{equation}
    In boosted version, the loss function is reformed into UIB style as:
    \begin{equation}
        \begin{split}
            \mathcal{L}' =  - \sum_{(u,p) \in \mathcal{I}} \text{ln}\sigma (s(u,p) - b_u) \\
                            - \alpha \sum_{(u,n) \notin \mathcal{I}} \text{ln}\sigma (b_u - s(u,n)).
        \end{split}
        \label{eq:blns}
    \end{equation}

    \item \textbf{Pointwise model}, i.e. the NCF~\cite{NCF}, which is a classical neural collaborative filtering model. The NCF uses the cross-entropy loss (pointwise approach) to optimize the model as equation~\ref{eq:ce}. Here, we only use the MLP version of NCF as our baseline and replace the cross-entropy loss with UIB loss to build our boosted model. Specifically, we directly replace the loss function with equation~\ref{eq:blns} in the boosted version of NCF.
    \begin{equation}
        \mathcal{L} = - \sum_{(u,x) \in \mathcal{T}} l(u,x)\text{ln}(s(u,p)) + (1-l(u,x))\text{ln}(1-s(u,n))
        \label{eq:ce}
    \end{equation}

    \item \textbf{Model with complicated loss function}, i.e. the SML~\cite{SML}, which is a S.O.T.A. metric learning model. 
    The loss function of SML not only makes sure the score of positive items is higher than that of negative, i.e. $\mathcal{L}_A$, but also keeps positive items away from negative items, i.e. $\mathcal{L}_B$. 
    Furthermore, it extends the traditional margin loss with the dynamically adaptive margins to mitigate the impact of bias. 
    The loss of SML can be characterized as:
    \begin{align}
        \mathcal{L}_A & = |s(u,n) - s(u,p)+m_u|_+ \\
        \mathcal{L}_B & = |s(n,p) - s(u,p)+n_u|_+ \\
        \mathcal{L}   & = \sum_{(u,p) \in \mathcal{I}} \sum_{(u,n) \notin \mathcal{I}}  
                          \mathcal{L}_A + \lambda \mathcal{L}_B + \gamma \mathcal{L}_{AM}
    \end{align}
    where $m_u$ and $n_u$ are learnable margin parameters, $\lambda$ and $\gamma$ are hyperparameters and the $\mathcal{L}_{AM}$ is the regularization on dynamical margins.

    Since the loss of SML contains multiple pairwise terms, various adaptations on SML with UIB can be selected, which also shows the flexibility of our approach to boost models with a complex loss function. 
    Here, we boost the SML only by replacing the main part $\mathcal{L}_A$ to:
    \begin{equation}
        \mathcal{L}'_A = |s(u,n) - b_u + m_u|_+ + \alpha |b_u - s(u,p) + m_u|_+
    \end{equation}

    \item \textbf{Model with complicated feature encoding}, i.e. the LightGCN~\cite{LightGCN}, is used to make sure our approach can work on advanced models.
    The LightGCN~\cite{LightGCN} is a state-of-the-art graph convolution network for the recommendation task. 
    Here, we only focus on the loss part and employ the simplified $\text{G}(\cdot)$ to represent the complicated feature encoding process with the graph convolution network. 
    LightGCN uses the same loss function as BPR~\cite{BPR} in equation~\ref{eq:lns}.
    To boost the LightGCN, we remain the score function $s(u,x)=\text{G}(u)^{\top} \text{G}(x)$ and only replace the loss function with equation~\ref{eq:blns}.
\end{itemize}

\subsection{Datasets}
In order to evaluate the performance of each model comprehensively, we select four publicly available datasets including different types and sizes. The detailed statistics of the datasets are shown in table~\ref{table:datasets}. 
\begin{itemize}
    \item Amazon Instant Video (AIV) is a video subset of Amazon Dataset benchmark\footnote{https://jmcauley.ucsd.edu/data/amazon/}, which contains product reviews and metadata from Amazon ~\cite{AIV}. We follow the 5-core, which promises that each user and item have 5 reviews at least.
    \item LastFM dataset~\cite{LastFM} contains music artist listening information from Last.fm online music system \footnote{https://grouplens.org/datasets/hetrec-2011/}.
    \item Movielens-1M (ML1M) \footnote{https://grouplens.org/datasets/movielens/} dataset~\cite{movielens} contains 1M anonymous movie ratings to describe users' preferences on movies. 
    \item Movielens-10M (ML10M) dataset is the large version of ML1M, which contains 10 million ratings of 10,677 movies by 72,000 users. We use this dataset to check whether our approach works well on the large dataset.
\end{itemize}

\begin{table}[h]
    \begin{center}
    \begin{tabular}{ l | r | r | r | r }
    \hline
    Dataset     
    & \textbf{AIV}
    & \textbf{LastFM}
    & \textbf{ML1M}
    & \textbf{ML10M}
    \\
    \hline
    \#User      &  5,130 &  1,877 &   6,028 &    69,878 \\
    \#Item      &  1,685 & 17,617 &   3,706 &    10,677 \\
    \#Train     & 26,866 & 89,047 & 988,129 & 9,860,298 \\
    \#Valid     &  5,130 &  1,877 &   6,028 &    69,878 \\
    \#Test      &  5,130 &  1,877 &   6,028 &    69,878 \\
    \hline
    \end{tabular}
\end{center}
\caption{Dataset statistics.}
\label{table:datasets}
\end{table}


\subsection{Evaluation Protocol}
We use Hit Ratio(Hit@K), Normalized Discounted Cumulative Gain(NDCG@K) and Mean Reciprocal Rank(MRR@K) to evaluate the models, where K is selected from classical settings \{1, 10\}. 
The higher value of all measures means the better performance. 
As Hit@1, NDCG@1 and MRR@1 are equivalent mathematically, we only report the Hit@1, Hit@10, NDCG@10 and MRR@10. 

All datasets are split by the popular One-Leave-Out strategy~\cite{BPR,NCF} as table~\ref{table:datasets}.
In the testing phase, models learned are asked to rank a given item list for each user.
As the space of negative sampling is extremely huge or unknown in the real world, for each positive sample in the test set, we fix the number of its relevant negative items as 100 negative samples~\cite{SML}.
Each experiment is independently repeated 5 times on the same candidates and the average performance is reported.


\subsection{Settings}
We use the PyTorch~\cite{PyTorch} to implement all models and use Adagrad~\cite{Adagrad} optimizer to learn all models.
To make the comparison fair, the dimension $d$ and batch size of all experiments are assigned as $32$ and $1024$, respectively.
For all boosted version, $M=32$ is used to balance classes as discussed in section~\ref{sec:class-balance}.
Grid search with early stop strategy on NDCG@10 of the validation dataset is adopted to determine the best hyperparameter configuration.
Each experiment runs 500 epochs for all datasets except ML10M, which is limited to 100 epochs due to its big size.
The detailed hyperparameter configuration table is reported in the appendix.

\subsection{Performance Boosting(RQ1)}
From the results shown in figure~\ref{fig::performance}, we make three observations:
(1) Comparing our boosted versions and the baselines on all datasets, our approach works well for all data sets to improves the model, even if in the large ML10M dataset. It confirms that our model successfully uses the UIB to improve the prediction performance on the recommendation task.  Specifically, compared with the baselines, our models achieve a consistent average improvement of 6.669\% on HIT@1, 1.579\% on HIT@10, 3.193\% on NDCG@10, 4.078\% on MRR@10 for the AIV dataset, 4.153\%, 1.013\%, 2.345\%, 2.909\% for the LastFM dataset, 5.075\%, 0.782\%, 2.001\%, 2.608\% for the ML1M dataset and 4.987\%, 0.2946\%, 2.341\%, 3.277\% for the ML10M dataset. 
Among them, the improvement on HIT@1 is the most impressive, reaching 5.22\% average on all datasets.
(2) Comparing the pointwise and pairwise models, our approach achieves a higher performance, concretely, 9.18\%, 0.68\%, 3.83\%, 5.34\% average improvements for pointwise approach and 5.81\%, 1.36\%, 3.10\%, 3.94\% for pairwise approach. It confirms that the UIB loss can be used to boost pointwise or pairwise based models dominated in the recommendation system. 
It is because our approach can help for learning inherent ranking property to boost the pointwise and improves training efficiency to boost the pairwise.
(3) Compared to the state-of-the-art model LightGCN~\cite{LightGCN} with complicated feature encoding, our boosted model LightGCN+UIB makes an average gain of 4.73\% on HIT@1, 0.93\% on HIT@10, 2.30\% on NDCG@10 and 2.97\% on MRR@10, which suggests our approach also works on the deep learning models.

\begin{table*}[h]
    \begin{center}
    \begin{tabular}{ l | c|c|c|c  ||  c|c|c|c}
    \hline
                    & \multicolumn{4}{c||}{\textbf{AIV}}             & \multicolumn{4}{c}{\textbf{LastFM}}          \\
    \hline
                        & HIT@1     & HIT@10    & NDCG@10   & MRR@10    & HIT@1     & HIT@10    & NDCG@10   & MRR@10    \\
    \hline
    BPR                 &   0.2216	& 0.5776	& 0.3848	& 0.3250    & 0.4717    & 0.7903	& 0.6336	& 0.5830    \\
    BPR+UIB(ours)       &   0.2452	& 0.5949	& 0.4063	& 0.3477    & 0.4907    & 0.7995	& 0.6493	& 0.6006    \\
    \hline
    NCF                 &   0.2322	& 0.6268	& 0.4149	& 0.3489    & 0.4827    & 0.8034	& 0.6445	& 0.5934    \\
    NCF+UIB(ours)       &   0.2608	& 0.6314	& 0.4333	& 0.3714    & 0.5205    & 0.8135	& 0.6727	& 0.6270    \\
    \hline
    SML                 &   0.2624	& 0.6796	& 0.4559	& 0.3861    & 0.5004    & 0.8237	& 0.6681	& 0.6175    \\
    SML+UIB(ours)       &   0.2591	& 0.6889	& 0.4582	& 0.3863    & 0.5077    & 0.8258	& 0.6711	& 0.6210    \\
    \hline
    LightGCN            &   0.2617	& 0.6732	& 0.4540	& 0.3855    & 0.5070    & 0.8141	& 0.6644	& 0.6160    \\
    LightGCN+UIB(ours)  &   0.2747	& 0.6814	& 0.4642	& 0.3964    & 0.5237    & 0.8253	& 0.6782	& 0.6307    \\
    \hline
    \textbf{Average Improvement} 
                        & \textbf{6.669\%} & \textbf{1.579\%} & \textbf{3.193\%} & \textbf{4.078\%} 
                        & \textbf{4.153\%} & \textbf{1.013\%} & \textbf{2.345\%} & \textbf{2.909\%}	\\
    \hline

    \hline
                    & \multicolumn{4}{c||}{\textbf{ML1M}}           & \multicolumn{4}{c}{\textbf{ML10M}}            \\
    \hline
                        & HIT@1     & HIT@10    & NDCG@10   & MRR@10    & HIT@1     & HIT@10    & NDCG@10   & MRR@10    \\
    \hline
    BPR                 & 0.3327    & 0.8135	& 0.5604	& 0.4807    & 0.5279    & 0.9380	& 0.7336	& 0.6680    \\
    BPR+UIB(ours)       & 0.3486    & 0.8205	& 0.5741	& 0.4964    & 0.5478    & 0.9418	& 0.7473	& 0.6846    \\
    \hline
    NCF                 & 0.3272    & 0.8161	& 0.5581	& 0.4770    & 0.5278    & 0.9419	& 0.7360	& 0.6697    \\
    NCF+UIB(ours)       & 0.3515    & 0.8222	& 0.5753	& 0.4975    & 0.5760    & 0.9416	& 0.7611	& 0.7029    \\
    \hline
    SML                 & 0.3013    & 0.7946	& 0.5319	& 0.4496    & 0.4806    & 0.9261	& 0.7030	& 0.6315    \\
    SML+UIB(ours)       & 0.3132    & 0.8021    & 0.5352    & 0.4512    & 0.4832    & 0.9283    & 0.7102    & 0.6411    \\
    \hline
    LightGCN            & 0.3329    & 0.8135	& 0.5613	& 0.4818    & 0.5182    & 0.9339    & 0.7254    & 0.6585    \\
    LightGCN+UIB(ours)  & 0.3467    & 0.8182	& 0.5717	& 0.4939    & 0.5519    & 0.9392    & 0.7476    & 0.6858    \\
    \hline
    \textbf{Average Improvement} 
                        & \textbf{5.075\%} & \textbf{0.7820\%} & \textbf{2.001\%} & \textbf{2.608\%} 
                        & \textbf{4.987\%} & \textbf{0.2946\%} & \textbf{2.341\%} & \textbf{3.277\%}	\\
    \hline
\end{tabular}
\end{center}
\caption{Performance comparison on four datasets.}
\label{fig::performance}
\end{table*}

\subsection{User Interest Boundary(RQ2)}
\label{sec:uib}

\begin{figure}
    \centering
    \includegraphics[width=0.97\columnwidth]{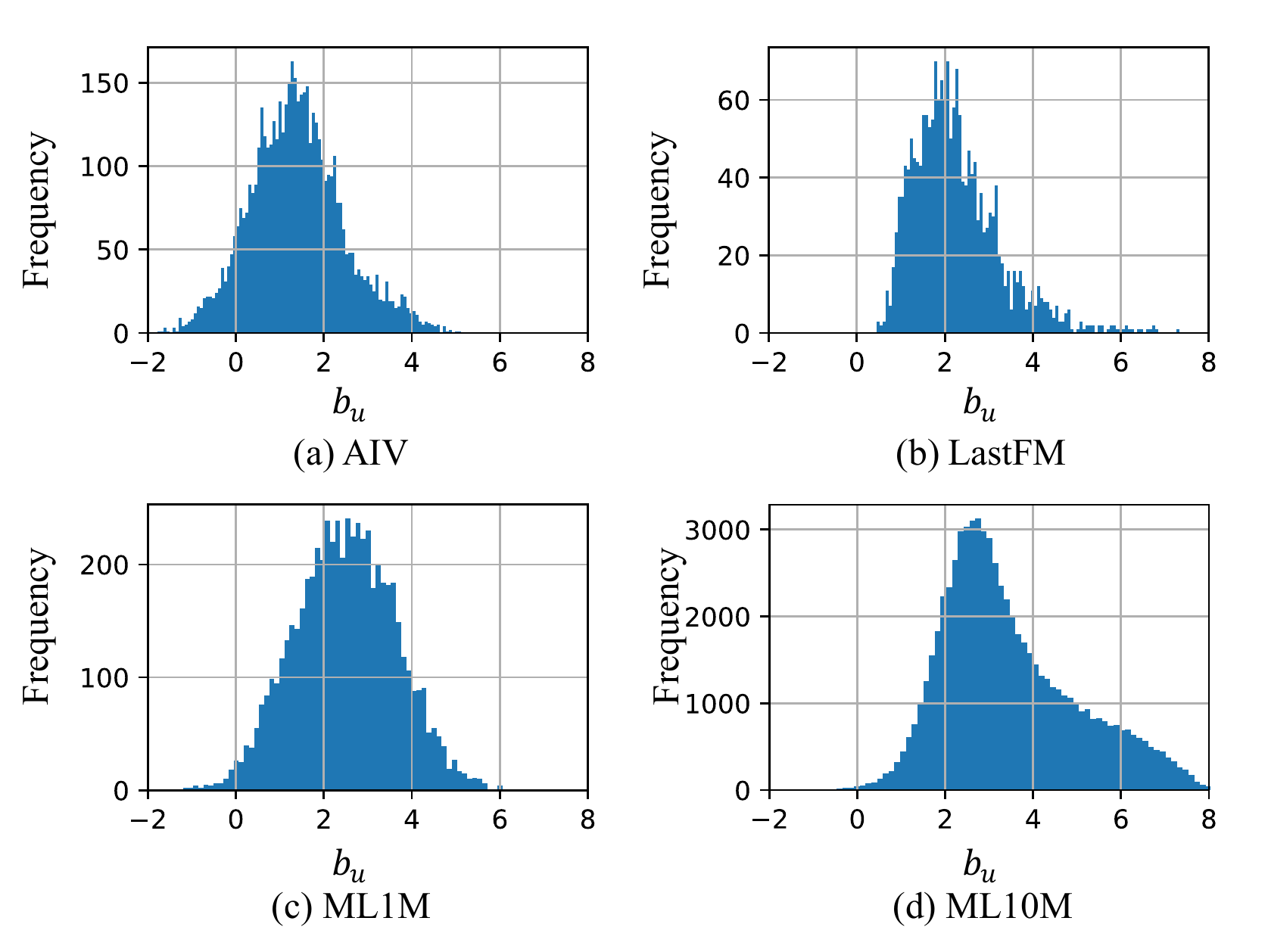}
    \caption{Boundary distribution on different datasets.}
    \label{fig:bu_distribution}
\end{figure}

\begin{figure}
    \centering
    \includegraphics[width=0.97\columnwidth]{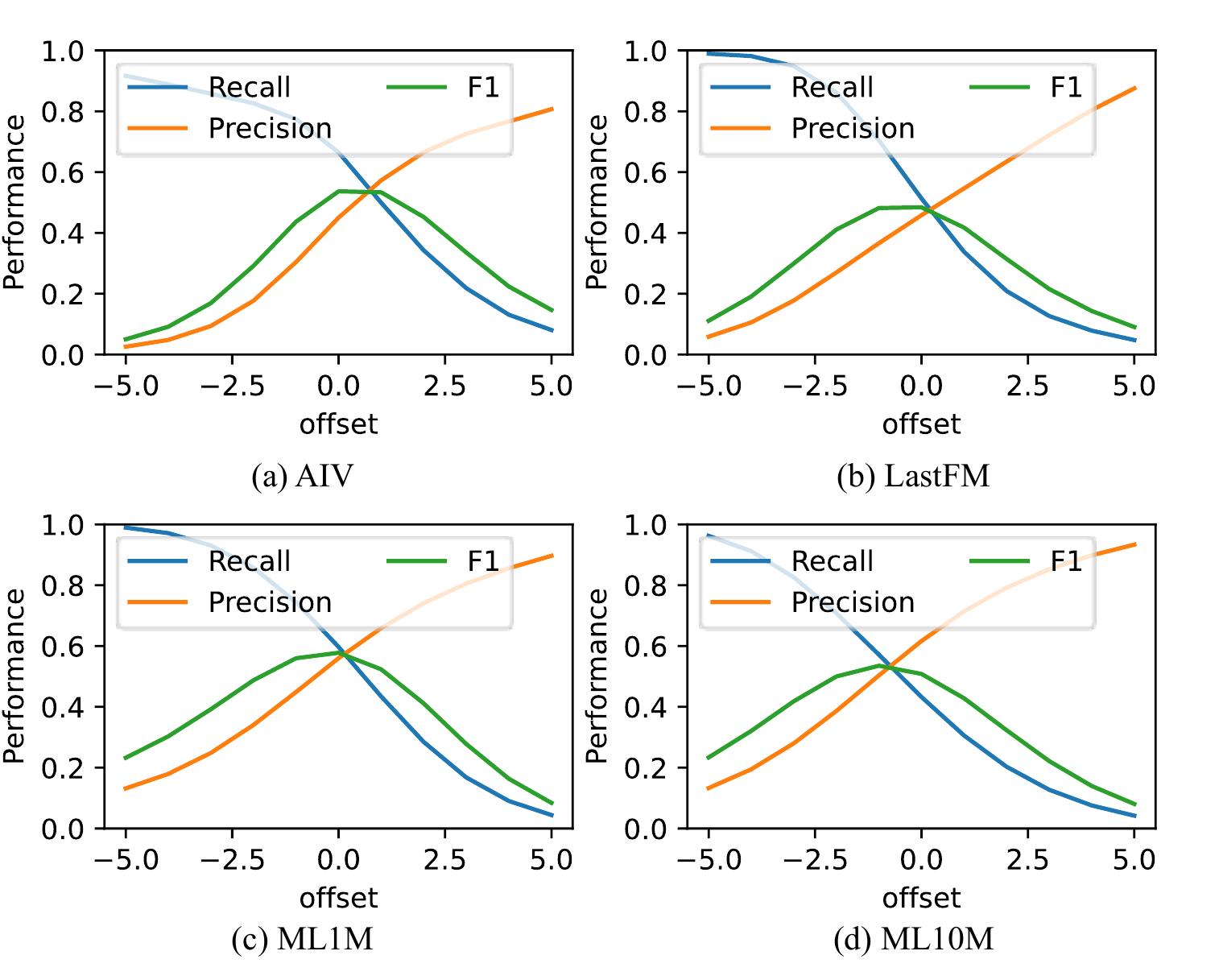}
    \caption{Measures with different offset on boundary.}
    \label{fig:bu_recall}
\end{figure}

To determine how well the boundary matches the user's interest scope, 
we essentially need to answer two sub-questions:
(1) Does the model match different boundaries for different users?
(2) Does the matched boundary is the best value?
To answer the first question, learned boundary distributions of NCF+UIB on four datasets are visualized in figure~\ref{fig:bu_distribution}. 
It confirms that our model matches different boundaries for different users in the form of different normal distributions.
To answer the second question, 
we add extra offsets \{-5,-4 $\dots$ 4,5\} on the $b_u$ and make predictions on all items for users.
Then the precision, recall, and F1 measures are reported to determine if the boundary learned is the best.
As the results shows in figure~\ref{fig:bu_recall}, 
(1) The boundary learned makes a competitive performance. Specifically, the F1 measure of 54\% in AIV, 48\% in LastFM, 58\% in ML1M, and 51\% in ML10M are achieved, which suggests the boundary is competent to match the user's interest scope.
(2) From all datasets, reducing offset consistently increases recall and reduces accuracy, and vice versa. When the offset is zero, F1 that considers both precision and recall performs best which shows that our method can learn the best boundary matching the user's interest scope. 
As discussed in section~\ref{sec:UIB}, the boundary is the best result of game between positive and negative samples, 
which can save computing by filtering out abundant obviously worthless items in the pre-ranking stage,
like 1674 of 1685 items (99.37\%) for AIV, 
17562 of 17617(99.69\%) fro LastFM,
3529 of 3706(95.24\%) for ML1M, 
10576 of 10677 (99.06\%) for ML10M.




\subsection{Hyperparameter Studies(RQ3)}
\label{sec:alpha}
\begin{figure}[h]
    \centering
    \includegraphics[width=0.97\columnwidth]{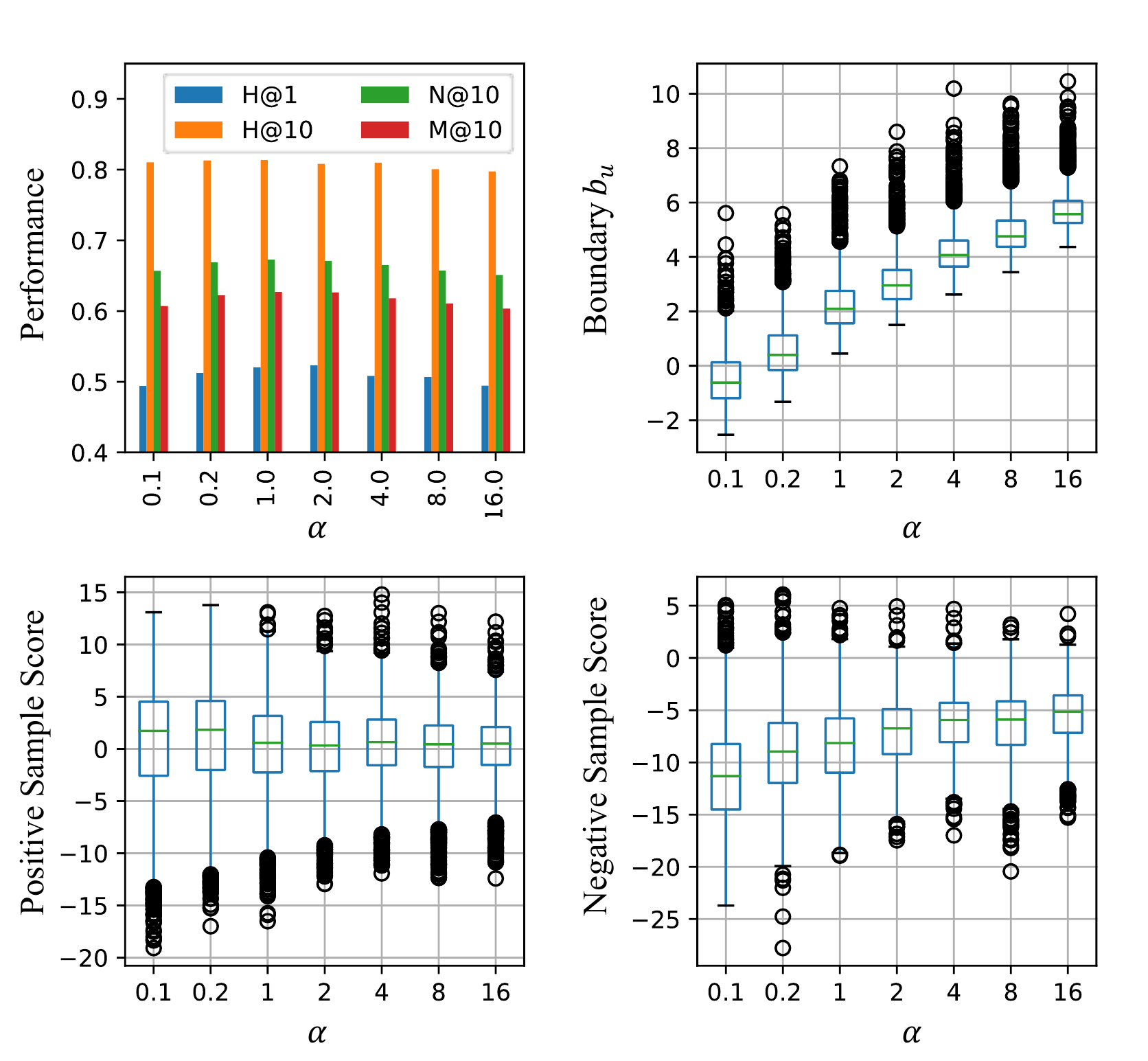}
    \caption{Model comparison with different $\alpha$ settings.}
    \label{fig:with_alpha}
\end{figure}

In our approach, only one hyperparameter $\alpha$ is introduced to balance the contributions of positive and negative samples.
To investigate how does the $\alpha$ affect the behaviour of our approach, experiments with various $\alpha$ $\in$ \{0.1,0.2,1,2,4,8,16\} settings are conducted based on the optimal experiment of NCF+UIB in LastFM datasets. We also provides the results on other datasets in the appendix.
%
%
Besides the performance and boundary distribution comparison, we also analyze the score distribution changes of positive and negative samples.
The experimental results in figure~\ref{fig:with_alpha} show that: 
(1) From the first cell of figure~\ref{fig:with_alpha}, it is confirmed that the $\alpha$ does affect the performance and here exists an optimal $\alpha$ to achieve the best performance. 
(2) Along with the growth of $\alpha$, boundary $b_u$ consistently increases, which suggests that the $\alpha$ is strongly relative with boundary distribution. 
This phenomenon confirms that increasing $\alpha$ to emphasize the negative loss part actually forces upward the boundary and affects both positive and negative sides at the same time.
(3) From the change of positive sample score distribution at the bottom-left, it is confirmed that the positive sample score space is compressed caused by the growth of $\alpha$.
(4) We also observe that the negative sample score distribution becomes compact at the bottom-right cell.
It is because the larger $\alpha$ also enlarges the loss of ``margin'', such as $\gamma$ in the MarginLoss $\text{max}(0, \Delta + \gamma)$, which pushes negative samples farther from the boundary and becomes compact.



\subsection{Training Efficiency (RQ4)}
\label{sec:analysis-efficiency}
To show the ability of our approach to improve training efficiency and alleviate the gradient vanishing problem that plague the traditional pairwise approach, 
experiments on BPR(the pairwise approach) and BPR+UIB(ours) are conducted to compare the rate of corrupted samples through epochs, i.e. the proportion of training samples that model classify incorrectly.
Usually, a high corrupted rate means that a higher proportion of training samples can provide gradient information to optmize model.

\begin{figure}
    \centering
    \includegraphics[width=0.6\columnwidth]{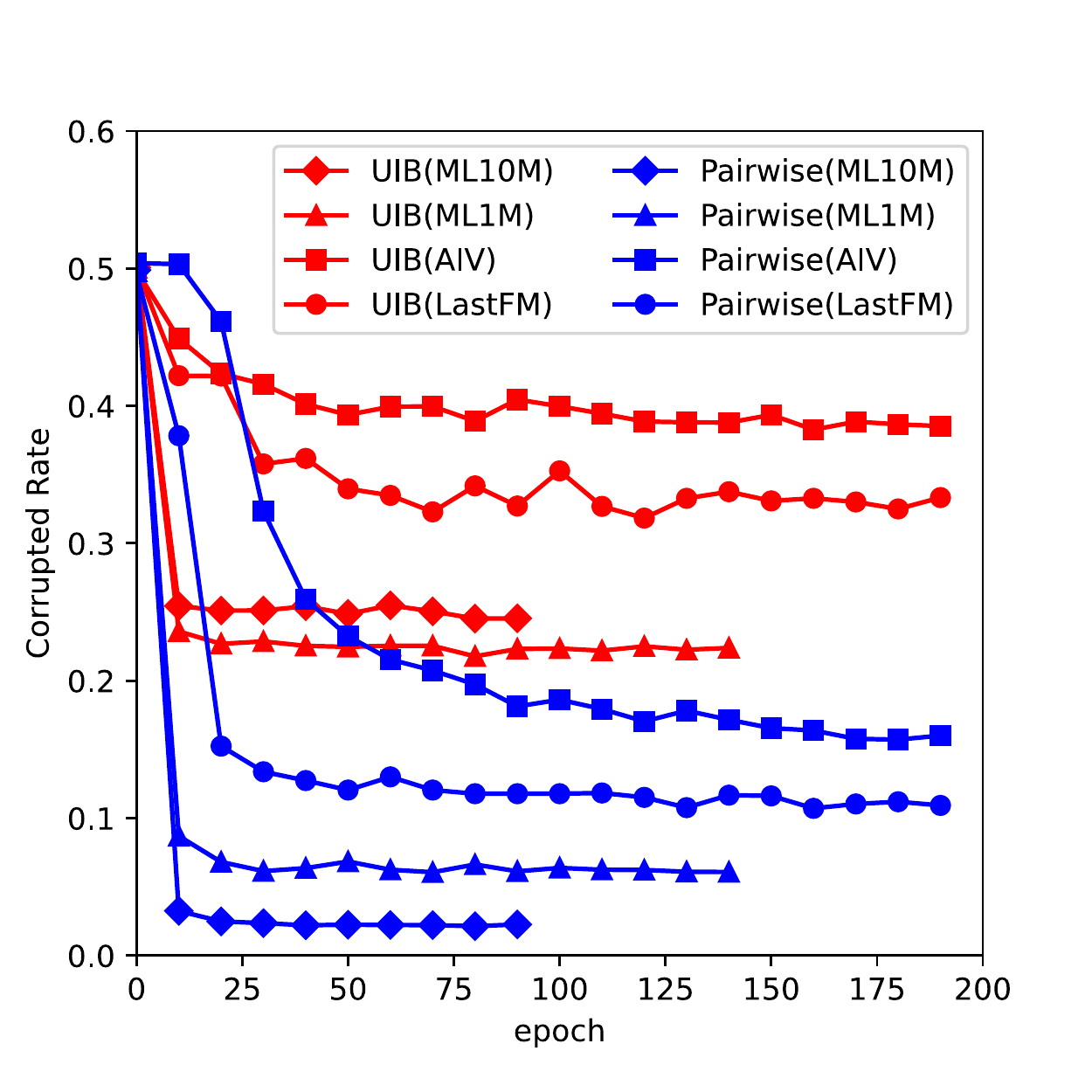}
    \caption{Training efficiency over epoch.}
    \Description{Training efficiency over epoch.}
    \label{fig:efficiency}
\end{figure}

As shown in figure~\ref{fig:efficiency}, the x-axis is the epoch through training, while the y-axis is the corrupted rate. 
The red is computed by our approach, while the blue is by the traditional pairwise approach.
From figure~\ref{fig:efficiency}, we can observe that the corrupted rate in our model is higher than that in BPR loss on all datasets. Consistent with previous literature, the training efficiency of the traditional pairwise model decreases obviously with the convergence, that is, the gradient vanishing. 
This leads to low training efficiency. Especially in ML10M and ML1M datasets, the proportion of training samples that can provide effective gradient is low after 10 epochs. 
Our approach maintains a certain effective gradient in training for all datasets, especially in the AIV dataset. 
As discussed in section~\ref{sec::training_eff}, this is because the boundary itself is a good hard sample to guide the training.






\section{Related Work}
This paper tries to combine and complement the two mainstream loss paradigms for the recommendation task.
As the pointwise and the pairwise approaches have their advantages and limitations~\cite{A-Case-Study}, several methods have been proposed to improve the loss function~\cite{DeepFM,yang2021top}.
\citet{nce} improved the recommendations based on the formation of user neighbourhoods.
\citet{DGR} proposed Wasserstein automatic coding framework to improve data sparsity and optimize uncertainty.
The pairwise approach is good at modeling the inferent ranking property but suffers the inflexible optimization~\cite{CircleLoss}.
\citet{PPNW} proposed a personalized pairwise weighting framework for BPR loss function, which makes the model overcome the limitations of BPR on cold start of items.
\citet{RLPP} proposed a model that can jointly learn the new representation of users and items in the embedded space, as well as the user's preference for item pairs.
\citet{SimpleX} considers loss function and negative sampling ratio equivalently and propose a unified CF model to incorporate both.
\citet{zhou2017learning} introduces a limitations to ensure the fact that the scoring of correct instances must be low enough to fulfill the translation.
Inspired by metric learning~\cite{CircleLoss}, several researchers try to employ the metric learning to optimize the recommendation model~\cite{kulis2013metric,Song_Nie_Han_Li_2017,CML,TransCF,SML}.
In addition, the problem of low training efficiency of paired method has also attracted much attention.
Uniform Sampling approaches are widely used in training collaborative filtering because of their simplicity and extensibility~\cite{NCF}. 
Advanced methods try to mine the hard negative samples to improve the training efficiency, including attribute-based~\cite{pairwise2014,AOBPR}, GAN-based~\cite{IRGAN,RNS-AS,AdvIR}, Cache-based~\cite{NSCaching,SRNS} and Random-walk-based methods~\cite{WalkRanker}.
\citet{ENMF} provides another way by directly using all samples in the negative sample space.


\section{Conclusion}
In this work, we innovatively introduce an auxiliary score $b_u$ for each user to represent the User Interest Boundary(UIB) and individually penalize samples that cross the boundary with pairwise paradigms.
In this way, our approach successfully achieves a hybrid loss of the pointwise and the pairwise to combine the advantages of both.
Specifically, it follows the pointwise in the whole loss expression while the pairwise inside each example.
Analytically, we show that 
our approach can provide a personalized decision boundary and significantly improve the training efficiency without any special sampling strategy.
Extensive results show that our approach achieves significant improvements on not only classical models of the pointwise or pairwise approaches, but also state-of-the-art models with complex loss function and complicated feature encoding.

\bibliographystyle{ACM-Reference-Format}
\bibliography{um}

\appendix
\section{Appendix}

\begin{figure*}[h]
    \centering
    \includegraphics[width=0.97\textwidth]{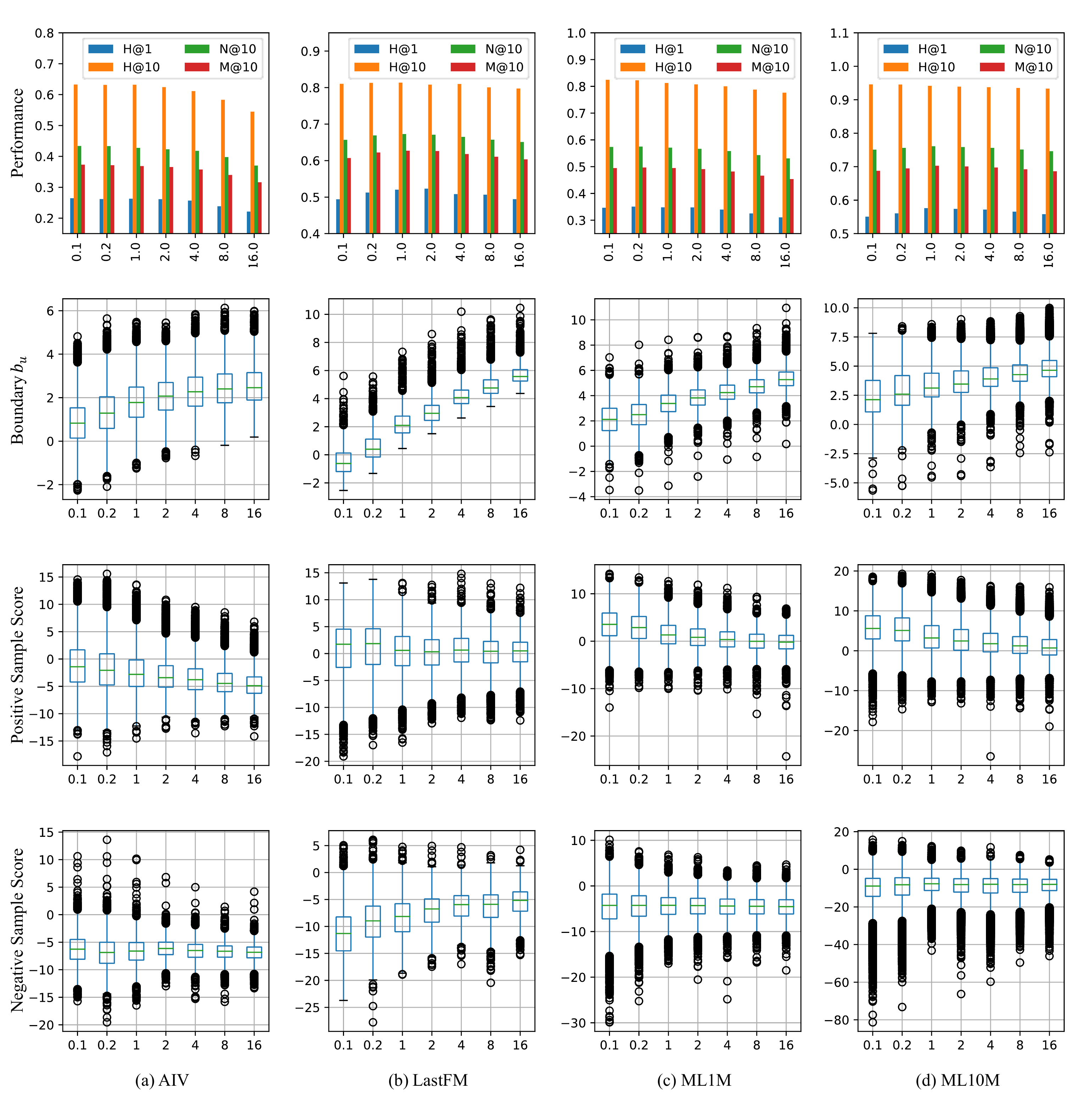}
    \caption{Model behaviours with different $\alpha$ settings.}
\end{figure*}

\begin{table*}
    \begin{center}
    \begin{tabular}{ l | c  | c | c | c | c}
    \hline
                &           & ML10M & ML1M  & AIV   & LastFM \\  
    \hline
    BPR         & $\eta$    & 1.0   & 1.0   & 3.0   & 1.0   \\
                & $\tau$    & 0.0   & 0.1   & 0.3   & 0.2   \\
    \hline
    BPR         & $\eta$    & 3.0   & 1.0   & 3.0   & 3.0   \\
    +UIB        & $\tau$    & 0.1   & 0.1   & 0.2   & 0.2   \\
                & $\alpha$  & 8.0   & 8.0   & 1.0   & 2.0   \\
    \hline
    NCF         & $\eta$    & 1.0   & 1.0   & 1.0   & 1.0   \\
                & $\tau$    & 0.1   & 0.1   & 0.4   & 0.3   \\
    \hline
    NCF         & $\eta$    & 1.0   & 1.0   & 1.0   & 1.0   \\
    +UIB        & $\tau$    & 0.1   & 0.1   & 0.4   & 0.4   \\
                & $\alpha$  & 8.0   & 8.0   & 0.1   & 8.0   \\
    \hline
    SML         & $\eta$    & 0.1   & 1.0   & 1.0   & 1.0   \\
                & $\tau$    & 0.0   & 0.0   & 0.0   & 0.0   \\
                & $\lambda$ & 0.3   & 0.3   & 0.3   & 0.3   \\
                & $\gamma$  & 64    & 128   & 256   & 128   \\
    \hline
    SML         & $\eta$    & 0.1   & 1.0   & 0.3   &  0.3  \\
    +UIB        & $\tau$    & 0.0   & 0.0   & 0.0   &  0.0  \\
                & $\lambda$ & 0.3   & 0.3   & 0.3   &  0.3  \\
                & $\gamma$  & 64    & 128   & 256   &  256  \\
                & $\alpha$  & 0.2   & 0.2   & 0.2   &  2.0  \\
    \hline
    LightGCN    & $\eta$    & 0.1   & 0.1   & 0.03  & 0.1   \\
                & $\tau$    & 0.0   & 0.0   & 0.0   & 0.0   \\
                & $\upsilon$& 1e-4  & 1e-4  & 1e-4  & 1e-4  \\
    \hline
    LightGCN    & $\eta$    & 0.3   & 0.3   & 0.03  & 0.1   \\
    +UIB        & $\tau$    & 0.0   & 0.0   & 0.0   & 0.0   \\
                & $\upsilon$& 1e-4  & 1e-4  & 1e-4  & 1e-4  \\
                & $\alpha$  & 8.0   & 8.0   & 8.0   & 0.2   \\
    \hline
    \end{tabular}
\end{center}
\caption{Model Hyperparameter Settings.}
\end{table*}

\end{document}